\documentstyle[twoside,fleqn,espcrc2]{article}

\newcommand{\Z}{{\sf Z \!\!\! Z}}

\newcommand{\AmS}{{\protect\the\textfont2
  A\kern-.1667em\lower.5ex\hbox{M}\kern-.125emS}}

\title{Perfect lattice actions with and without chiral symmetry}

\author{W. Bietenholz \address{Center for Theoretical Physics, Laboratory
of Nuclear Science and Department of Physics \\
Massachusetts Institute of Technology (MIT), Cambridge, MA 02139, USA}
and U.-J. Wiese$^{{\rm ~a}}$
\thanks{Based on a poster presented by WB and a talk given by UJW. This work
is supported in part by funds provided by the U.S. Department of Energy
(D.O.E.) under cooperative research agreement DE-FC02-94ER40818.}}

\begin{document}

\begin{abstract}

We use perturbation theory to construct perfect lattice actions for fermions
and gauge fields by blocking directly from the continuum. When one uses a
renormalization group transformation that preserves chiral symmetry the
resulting lattice action for massless fermions is chirally symmetric but
nonlocal. When the renormalization group transformation breaks chiral symmetry
the lattice action becomes local but chiral symmetry is explicitly broken.
In particular,
starting with a chiral gauge theory in the continuum one either obtains
a lattice theory which is gauge invariant but nonlocal, or a local theory with
explicitly broken gauge invariance. In both cases the spectrum of the lattice
theory is identical with the one of the continuum and the anomaly is correctly
reproduced. We also apply our techniques to vector-like theories. In
particular we propose a new renormalization group transformation for QCD and we
optimize its parameters for locality of the perfect action.

\end{abstract}

\maketitle

\section{Introduction}

When a field theory is regularized on the lattice it is unavoidable that
some symmetries of the continuum get broken explicitly. For example,
one gives up Poincar\'{e} symmetry and expects that it is restored in the
continuum limit. Chiral symmetry poses subtle problems in lattice theories
due to the presence of anomalies. Based on mild assumptions the
Nielsen-Ninomiya theorem implies that a chirally invariant lattice theory with
a local action leads to fermion doubling. Doubling can be avoided by breaking
chiral symmetry explicitly. In a vector-like theory such as QCD one expects
that
chiral symmetry is restored in the continuum limit. In a chiral gauge theory,
on the other hand, breaking chiral symmetry destroys gauge invariance. On the
lattice one usually tries to keep gauge invariance intact. If one breaks it
explicitly, restoring it in the continuum limit requires fine tuning of
counter term couplings which is complicated in a numerical simulation.
An alternative is to give up locality. Also this causes problems in
simulations.

Perfect actions are by definition free of cut-off effects. They are located
on a renormalized trajectory that intersects the critical surface in a fixed
point of a renormalization group transformation
\cite{Kog74}. Still, a perfect action is a
lattice object, and for example Poincar\'{e} symmetry can not be manifest.
However, the symmetry violation does not affect the spectrum of the theory,
which is identical with the one of the continuum. This is because the symmetry
breaking is entirely due to the renormalization group transformation.

Here we construct perfect actions for fermions and gauge fields. Like
Poincar\'{e} symmetry the chiral symmetry of the fermions may not be manifest
in the action, but it is still present in the spectrum of the theory. This is
the case when one uses a renormalization group transformation that breaks
chiral symmetry explicitly. The resulting fixed point action is local. In
a chiral gauge theory gauge invariance is then not manifest in
the action, although it is still present in the spectrum. Alternatively one
may choose a renormalization group transformation that preserves chiral
symmetry. Then gauge invariance in a chiral gauge theory is manifest in the
action, but the lattice theory becomes nonlocal. Again, by construction the
nonlocality does not affect the physical content of the theory. Here
nonlocality arises naturally by integrating out degrees of freedom at short
distances.

We use a renormalization group transformation that blocks
directly from the continuum \cite{Wil76}.
Perfect actions for fermions and gauge fields
are derived in perturbation theory and the perfect fermion-gauge vertex is
obtained to leading order in the gauge coupling. This leads to a perturbative
definition of lattice chiral gauge theories. At present it is unclear how our
construction may be extended nonperturbatively. Our results also apply to
vector-like theories. For example, we show explicitly that the correct axial
anomaly is reproduced in the Schwinger model \cite{Bie95}.

The analytic calculations
presented here are a step towards the perfect action of full QCD. Then one
demands a very local action, which implies non-manifest chiral symmetry.
We optimize the parameters of our renormalization group transformation such
that the perfect action becomes ultralocal for free fermions in one dimension
and for free gauge fields in two dimensions. We also propose a new block factor
2 renormalization group transformation for QCD that reduces to blocking from
the continuum for weak fields. This transformation shall be used later in a
nonperturbative determination of a perfect QCD action.

We follow the concept to approximate perfect actions proposed
in \cite{Has94}. Work on the perfect action for pure $SU(3)$
gauge theory was presented in \cite{DeG95}. For a different approach
to damp cut-off artifacts in the lattice action, which has attracted
much attention at this conference, see \cite{Lep93}.

\section{Blocking free fermions from the continuum}

We consider a fermion field $\bar\psi, \psi$ in the continuum and
construct a lattice fermion field by averaging the continuum field over unit
hypercubes $c_x$ centered at the points $x$ of a $d$-dimensional Euclidean
lattice
\begin{equation}
\bar\Psi_x = \int_{c_x} d^dy \ \bar\psi(y),  \,\,\,
\Psi_x = \int_{c_x} d^dy \ \psi(y).
\end{equation}
In momentum space the lattice fermion field is given by
\begin{eqnarray}
\hspace{-7mm}
&&\Psi(p)=\sum_{l \in \Z^d} \psi(p + 2 \pi l) \Pi(p + 2 \pi l), \nonumber \\
\hspace{-7mm}
&&\Pi(p)=\prod_{\mu = 1}^d \frac{\hat p_\mu}{p_\mu}, \,\,\,
\hat p_\mu = 2 \sin(p_\mu/2).
\end{eqnarray}
Note that the momentum of the lattice field is restricted to the Brillouin
zone $B = ]-\pi,\pi]^d$. Due to the summation over the integer
vector $l$ the lattice field is periodic over $B$. Now the continuum degrees
of freedom are integrated and one arrives at an effective action for the
lattice variables
\begin{eqnarray}
\label{smearedRGT}
\hspace{-7mm}
&&\exp(- S[\bar\Psi,\Psi])=\int {\cal D}\bar\psi {\cal D}\psi
{\cal D}\bar\eta {\cal D}\eta \nonumber \\
\hspace{-7mm}
&&\times\exp\Big\{- \frac{1}{(2 \pi)^d} \int d^dp \
\bar\psi(-p) \delta^f(p)^{-1} \psi(p)\Big\} \nonumber \\
\hspace{-7mm}
&&\times\exp\Big\{\frac{1}{(2 \pi)^d} \int_B d^dp \Big[a \bar\eta(-p) \eta(p)
\nonumber \\
\hspace{-7mm}
&&\times[\bar\Psi(-p) - \sum_{l \in \Z^d} \bar\psi(- p - 2 \pi l)
\Pi(p + 2 \pi l)] \eta(p) \nonumber \\
\hspace{-7mm}
&&+\bar\eta(-p)[\Psi(p) - \sum_{l \in \Z^d} \psi(p + 2 \pi l)
\Pi(p + 2 \pi l)]\Big]\Big\}.
\end{eqnarray}
Here $s[\bar\psi,\psi]$ is the continuum action for free fermions. In momentum
space it takes the form
\begin{eqnarray}
\hspace{-7mm}
&&s[\bar\psi,\psi] = \frac{1}{(2 \pi)^d} \int d^dp \
\bar\psi(-p) \delta^f(p)^{-1} \psi(p), \nonumber \\
\hspace{-7mm}
&&\delta^f(p) = (i \gamma_\mu p_\mu  + m)^{-1},
\end{eqnarray}
where $\delta^f(p)$ is the free fermion propagator in the continuum.
The fields $\bar\eta$ and $\eta$ are auxiliary lattice Grassmann variables.
For $a = 0$ this is a $\delta$-function renormalization group transformation
which is chirally invariant. Its fixed point (at $m=0$) describes free lattice
chiral fermions \cite{Wie93}. The corresponding fixed point action, however,
is nonlocal. This is consistent with the Nielsen-Ninomiya theorem.
For $a>0$ the renormalization group transformation breaks chiral symmetry
explicitly. However, as we will show, the chiral breaking introduced in the
renormalization group step does not affect the spectrum of the theory.

Performing the integrations in eq.(\ref{smearedRGT}) results in
\begin{eqnarray}
\label{trajectory}
S[\bar\Psi,\Psi]&=&\frac{1}{(2 \pi)^d} \int_B d^dp \
\bar\Psi(-p) \Delta^f(p)^{-1} \Psi(p), \nonumber \\
\Delta^f(p)&=&\sum_{l \in \Z^d} [i \gamma_\mu (p_\mu + 2 \pi l_\mu) + m]^{-1}
\nonumber \\
&\times&\Pi(p + 2 \pi l)^2 + a.
\end{eqnarray}
Here $\Delta^f(p)$ is the lattice fermion propagator.
The lattice action describes a complete renormalized trajectory parametrized
by the mass $m$. At $m=0$ the trajectory intersects the critical surface in
the fixed point.

Performing the Gaussian integral is equivalent to solving a classical equation
of motion for the continuum quark field. The solution of this equation is
\begin{eqnarray}
\label{psiequation}
\hspace{-7mm}
&&\bar\psi_c(-p)=\bar\Psi(-p) \Delta^f(p)^{-1} \Pi(p) \delta^f(p), \nonumber \\
\hspace{-7mm}
&&\psi_c(p)=\delta^f(p) \Pi(p) \Delta^f(p)^{-1} \Psi(p).
\end{eqnarray}
These relations between the continuum and lattice fermion fields will be useful
when we determine the perfect vertex function.

The introduction of nonzero $a$ leads to a local perfect action with
explicitly broken chiral symmetry even at $m = 0$. We now vary $a$ such that
the action for 1-dimensional configurations becomes ultralocal. For $d=1$
the fermion propagator takes the form
\begin{eqnarray}
\hspace{-7mm}
&&\Delta^f(p)=\frac{1}{m} - \frac{2}{m^2} [\coth(m/2) - i
\cot(p/2)]^{-1}\!+\!a.
\nonumber \\ \hspace{-7mm} && \,
\end{eqnarray}
Choosing $a = (\hat m - m)/m^2$, where $\hat m = e^m - 1$, the propagator
reduces to
\begin{equation}
\Delta^f(p) = \Big(\frac{\hat m}{m}\Big)^2
(i \sin p + \hat m + \frac{1}{2} \hat p^2)^{-1}.
\end{equation}
This corresponds to the standard nearest neighbor Wilson fermion action.
It turns out that the perfect action is still very local even in $d = 4$
\cite{BW95}.

Now we consider the spectrum of the fermionic lattice theory. The fermionic
operator with definite spatial momentum $\vec{p} \in ]- \pi,\pi]^{d-1}$,
\begin{equation}
\Psi(\vec{p})_{x_d} = \frac{1}{2 \pi} \int_{-\pi}^\pi dp_d \ \Psi(p)
\exp(i p_d x_d),
\end{equation}
creates a fermion at Euclidean time $x_d$. Its correlation function
is given by
\begin{eqnarray}
\hspace{-7mm}
&&\langle\bar\Psi(- \vec{p})_0 \Psi(\vec{p})_{x_d}\rangle= \nonumber \\
\hspace{-7mm}
&&\frac{1}{2 \pi} \int_{-\infty}^\infty dp_d  \sum_{\vec{l} \in \Z^{d-1}}
\frac{m}{(\vec{p} + 2 \pi \vec{l})^2 + p_d^2 + m^2} \nonumber \\
\hspace{-7mm}
&&\times\prod_{j=1}^{d-1} \Big(\frac{\hat p_j}{p_j + 2 \pi l_j}\Big)^2
\Big(\frac{\hat p_d}{p_d}\Big)^2 \exp(i p_d x_d) + a \delta_{x_d,0}=
\nonumber \\
\hspace{-7mm}
&&\sum_{\vec{l} \in \Z^{d-1}} C(\vec{p} + 2 \pi \vec{l})
\exp(- E(\vec{p} + 2 \pi \vec{l}) x_d) + a \delta_{x_d,0}.
\nonumber \\ \hspace{-7mm} && \,
\end{eqnarray}
The sum over spatial $\vec{l}$ corresponds to infinitely many poles of the
integrand, and thus to infinitely many states that contribute an exponential
to the 2-point function. The energies of these states are given by the location
of the poles, i.e. $E(\vec{p} + 2 \pi \vec{l}) = - i p_d$ with
\begin{equation}
E(\vec{p} + 2 \pi \vec{l})^2 = - p_d^2 = (\vec{p} + 2 \pi \vec{l})^2 + m^2.
\end{equation}
This is exactly the energy of a particle with mass $m$ and momentum
$\vec{p} + 2 \pi \vec{l}$, i.e. the spectrum of the lattice theory is
identical with the one of the continuum --- cut-off effects are completely
eliminated. Therefore the continuous Poincar\'{e} symmetry is restored in
the spectrum.

\section{Blocking free gauge fields from the continuum}

In this section we apply the technique of blocking from the continuum to free
gauge fields $a_\mu$. We define a noncompact lattice gauge field
\begin{eqnarray}
A_{\mu,x}&=&\int_{c_{x-\hat\mu/2}} \!\!\!\! d^dy \
(1 + y_\mu - x_\mu) a_\mu(y) \nonumber \\
&+&\int_{c_{x+\hat\mu/2}} \!\!\!\! d^dy \ (1 - y_\mu + x_\mu) a_\mu(y),
\end{eqnarray}
where $x$ now refers to the center of the link, i.e. $x_\mu$ is a half-integer.
This construction is gauge covariant, i.e. when one performs
a gauge transformation $^\varphi\! a_\mu(x) = a_\mu(x) - \partial_\mu
\varphi(x)$ in the continuum, it induces a lattice gauge transformation
$^\Phi\! A_{\mu,x} =
A_{\mu,x} - \Phi_{x+\hat\mu/2} + \Phi_{x-\hat\mu/2}$. The
lattice gauge transformation is given by
$\Phi_x = \int_{c_x} d^dy \ \varphi(y)$. In momentum space the lattice
gauge field reads
\begin{eqnarray}
\hspace{-7mm}
&&A_\mu(p)=\sum_{l \in \Z^d} a_\mu(p + 2 \pi l) \Pi_\mu(p + 2 \pi l)
(-1)^{l_\mu}, \nonumber \\
\hspace{-7mm}
&&\Pi_\mu(p)=\frac{\hat p_\mu}{p_\mu} \Pi(p).
\end{eqnarray}
Because the link centers have a half-integer component $x_\mu$, the gauge field
is antiperiodic over the Brillouin zone in the $\mu$-direction. For
technical reasons we work in the Landau gauge $\sum_\mu p_\mu a_\mu(p) = 0$.
Then the continuum action takes the form
\begin{eqnarray}
\hspace{-7mm}
&&s[a]=\frac{1}{(2 \pi)^d} \int d^dp \ \frac{1}{2}
a_\mu(-p) \delta^g(p)^{-1} a_\mu(p), \nonumber \\
\hspace{-7mm}
&&\delta^g(p)=p^{-2},
\end{eqnarray}
where $\delta^g(p)$ is the free gluon propagator in the Landau gauge. As in
the fermionic case we want to optimize the renormalization group
transformation for locality of the fixed point action. To this end we
introduce an auxiliary vector field $D_\mu$, analogous to the auxiliary
fermion field $\eta$. For $D_\mu$ we introduce a smearing term
$\frac{1}{2} D_\mu (\alpha + \gamma \hat p_\mu^2) D_\mu$.
This allows us to optimize the
parameters in the renormalization group transformation such that the fixed
point action becomes ultralocal for 2-d configurations. The parameter $\alpha$
alone is not sufficient for this purpose. This renormalization group
transformation takes the form
\begin{eqnarray}
\label{gaugeRGT}
\hspace{-7mm}
&&\exp(- S[A])=\int {\cal D}a {\cal D}D \nonumber \\
\hspace{-7mm}
&&\times\exp\Big\{- \frac{1}{(2 \pi)^d}
\int d^dp \Big[\frac{1}{2}
a_\mu(-p) \delta^g(p)^{-1} a_\mu(p) \nonumber \\
\hspace{-7mm}
&&- i D_\mu(-p) a_\mu(p) \Pi_\mu(p)\Big] \Big\} \nonumber \\
\hspace{-7mm}
&& \times \exp \Big\{ - \frac{1}{(2 \pi)^d} \int_B d^dp \Big[
i D_\mu (-p) A_\mu (p) \nonumber \\ \hspace{-7mm} &&
+ \frac{1}{2}
D_\mu(-p) (\alpha + \gamma \hat p_\mu^2) D_\mu(p) \Big]\Big\}.
\end{eqnarray}
Performing the Gaussian integral for $a_\mu$ is again equivalent to
solving a classical equation of motion which yields
\begin{equation}
\label{aequation}
a_{\mu c}(p) = \delta^g(p) \Pi_\mu(p) \Delta_\mu^g(p)^{-1} A_\mu(p).
\end{equation}
Since $a_\mu$ is in the Landau gauge, using $p_\mu \Pi_\mu(p) =
\hat p_\mu \Pi(p)$ one obtains
\begin{equation}
\label{gaugecondition}
\sum_\mu \hat p_\mu \Delta_\mu^g(p)^{-1} A_\mu(p) = 0.
\end{equation}
We call this the `fixed point lattice Landau gauge' condition.
The fixed point action in this gauge takes the form
\begin{eqnarray}
\label{LandauFPaction} \hspace{-7mm} &&
S[A] = \frac{1}{(2 \pi)^d}
\int_B d^dp \ \frac{1}{2}
A_\mu(-p) \Delta_\mu^g(p)^{-1}
A_\mu(p), \nonumber \\ \hspace{-7mm} &&
\Delta^g_\mu(p) = \sum_{l \in \Z^d} (p + 2 \pi l)^{-2}
\Pi_\mu(p + 2 \pi l)^2 \nonumber \\ \hspace{-7mm}
&& \qquad \quad + \alpha + \gamma \hat p_\mu^2.
\end{eqnarray}
Here $\Delta^g_\mu(p)$ is the lattice gauge field propagator in the fixed point
lattice Landau gauge.

Again we are looking for the optimally local fixed point action. We turn to
$d=2$ and work with field strength variables $F$ that live on the
plaquettes. After some manipulations one obtains
\begin{eqnarray}
S[F]&=&\frac{1}{(2 \pi)^2} \int_B d^2p \ \frac{1}{2}
F(-p) \rho(p) F(p), \nonumber \\
\rho(p)^{-1}&=&\hat p_1^2 \Delta_2^g(p) + \hat p_2^2 \Delta_1^g(p) \nonumber \\
&=&(1 - \frac{1}{6} \hat p_1^2)(1 - \frac{1}{6} \hat p_2^2)
\nonumber \\
&+&\alpha(\hat p_1^2 + \hat p_2^2) + 2 \gamma \hat p_1^2 \hat p_2^2.
\end{eqnarray}
For $\alpha = 1/6$ and $\gamma = - 1/72$ one finds $\rho(p) = 1$. This
corresponds to the standard plaquette action which is ultralocal. Hence in
$d = 2$ the optimal choice for the smearing term is
$\frac{1}{12} D_\mu (1 - \frac{1}{12} \hat p_\mu^2) D_\mu$.
We use the same parameters in $d = 4$
and still find a very local action \cite{BW95}.

\section{The perfect fermion-gauge vertex}

Now we switch on interactions between fermions and $SU(N)$ gauge fields
to leading order in the gauge coupling $e$. The continuum fermion field is
denoted by
$\bar\psi^i$, $\psi^i$ with $i \in \{1,2,...,N\}$ and the continuum
gauge field is denoted by $a^a_\mu$ with $a \in \{1,2,...,N^2-1\}$.
In momentum space the vertex of a left-handed charged fermion coupled to the
gauge field takes the form
\begin{eqnarray}
\hspace{-7mm}
&&v[\bar\psi,\psi,a] = \frac{e}{(2 \pi)^{2d}} \int d^dp \ d^dq \nonumber \\
\hspace{-7mm}
&&\times\bar\psi^i(-p) i \gamma_\mu \frac{1}{2}(1 - \gamma_5)
a^a_\mu(p-q) \lambda^a_{ij} \psi^j(q).
\end{eqnarray}
Again we integrate out the continuum fields and derive the perfect
vertex of the lattice theory. Before we can do so
we need to define the renormalization group transformation up to $O(e)$. This
can be done in many ways, and we do not want to make a specific choice yet. In
general, for the fermions one can write
\begin{eqnarray}
\label{kernel}
\hspace{-7mm}
&&\Psi^i(p)=\sum_{l \in \Z^d} \psi^i(p + 2 \pi l) \Pi(p + 2 \pi l)
\nonumber \\
\hspace{-7mm}
&&+ e \sum_{l \in \Z^d} \frac{1}{(2 \pi)^d} \int d^dq \
K_\mu(p + 2 \pi l,q + 2 \pi l) \nonumber \\
\hspace{-7mm}
&&\times a^a_\mu(p-q) \lambda^a_{ij}
\psi^j(q + 2 \pi l).
\end{eqnarray}
Here $K_\mu(p,q)$ is a regular kernel that specifies the renormalization group
transformation. It describes how the continuum quark field in a hypercube
$c_x$ is transported to the lattice point $x$.

Compared to the free fermion calculation of section 2 we now have an extra
term
\begin{eqnarray}
\hspace{-7mm}
&&T[\bar\psi,\psi,a]=\frac{e}{(2 \pi)^{2d}} \int d^dp \ d^dq \nonumber \\
\hspace{-7mm}
&&\times[\bar\eta^i(-p) K_\mu(p,q) a^a_\mu(p-q) \lambda^a_{ij} \psi^j(q)
\nonumber \\
\hspace{-7mm}
&&-\bar\psi^i(-p) K_\mu(-q,-p) a^a_\mu(p-q) \lambda^a_{ij} \eta^j(q)]
\end{eqnarray}
in the renormalization group transformation and hence in the exponential of
eq.(\ref{smearedRGT}). The additional terms $v[\bar\psi,\psi,a]$ and
$T[\bar\psi,\psi,a]$ modify the exponent $E[\bar\psi,\psi] + E[a]$ of the
free theory to $E[\bar\psi,\psi,a] = E[\bar\psi,\psi] + E[a] +
v[\bar\psi,\psi,a] + T[\bar\psi,\psi,a]$. Since the
additional terms are $O(e)$ they change the equations of motion (and hence
their solutions) to $\bar\psi_c + \delta\bar\psi_c$ and $\psi_c +
\delta\psi_c$ and $a_c + \delta a_c$, where $\delta\bar\psi_c$, $\delta\psi_c$
and $\delta a_c$ are of $O(e)$.
Inserting the new solution into the new exponent yields
\begin{eqnarray}
\hspace{-7mm}
&&E[\bar\psi_c + \delta\bar\psi_c,\psi_c + \delta\psi_c,a_c + \delta a_c]
+ O(e^2) = \nonumber \\
\hspace{-7mm}
&&E[\bar\psi_c,\psi_c] + E[a_c] + v[\bar\psi_c,\psi_c,a_c] \nonumber \\
\hspace{-7mm}
&&+ T[\bar\psi_c,\psi_c,a_c] + O(e^2)= \nonumber \\
\hspace{-7mm}
&&S[\bar\Psi,\Psi] + S[A] + V[\bar\Psi,\Psi,A] + O(e^2).
\end{eqnarray}
The essential observation is that $\bar\psi_c$, $\psi_c$ and $a_c$ are
solutions
of classical equations of motion, i.e. extrema of the corresponding exponents.
This implies $E[\bar\psi_c + \delta\bar\psi_c,\psi_c + \delta\psi_c] =
E[\bar\psi_c,\psi_c] + O(e^2)$ and $E[a_c + \delta a_c] = E[a_c]
+ O(e^2)$. Therefore we read off the perfect vertex
\begin{eqnarray}
\hspace{-7mm}
&&V[\bar\Psi,\Psi,A]=v[\bar\psi_c,\psi_c,a_c] + T[\bar\psi_c,\psi_c,a_c]=
\nonumber \\
\hspace{-7mm}
&&\frac{1}{(2 \pi)^{2d}} \int_{B^2} d^dp \ d^dq \
\bar\Psi^i(-p) e V_\mu(p,q) \nonumber \\
\hspace{-7mm}
&&\times A^a_\mu(p-q) \lambda^a_{ij} \Psi^j(q).
\end{eqnarray}
Using eqs.(\ref{psiequation}) and (\ref{aequation}) one identifies the
vertex function as
\begin{eqnarray}
\label{vertex}
\hspace{-7mm}
&&V_\mu(p,q)=\Delta^f(p)^{-1} \Delta^g_\mu(p-q)^{-1} \nonumber \\
\hspace{-7mm}
&&\times\sum_{l,m \in \Z^d} \delta^g(p + 2 \pi l - q - 2 \pi m) \nonumber \\
\hspace{-7mm}
&&\times\Pi_\mu(p + 2 \pi l - q - 2 \pi m) (-1)^{l_\mu+m_\mu} \nonumber \\
\hspace{-7mm}
&&\times[\delta^f(p + 2 \pi l) i \gamma_\mu \frac{1}{2}(1 - \gamma_5)
\delta^f(q + 2 \pi m) \nonumber \\
\hspace{-7mm}
&&\times\Pi(p + 2 \pi l) \Pi(q + 2 \pi m) \nonumber \\
\hspace{-7mm}
&&+K_\mu(p + 2 \pi l,q + 2 \pi m) \delta^f(q + 2 \pi m) \Pi(q + 2 \pi m)
\nonumber \\
\hspace{-7mm}
&&-K_\mu(-q - 2 \pi m,-p - 2 \pi l) \nonumber \\
\hspace{-7mm}
&&\times\delta^f(p + 2 \pi l) \Pi(p + 2 \pi l)]\Delta^f(q)^{-1}.
\end{eqnarray}
This expression is in the fixed point lattice Landau gauge.
It contains products of up to six $\gamma$-matrices. These
may be decomposed into 1, $\gamma_\mu$, $\sigma_{\mu\nu}$, $\gamma_5$ and
$\gamma_\mu \gamma_5$. The last two do not occur in QCD because then parity is
conserved. The $\sigma_{\mu\nu}$-term represents a perfect `clover' action.

When the renormalization group transformation is chirally symmetric (i.e. when
$a=0$) the lattice propagator $\Delta^f$ is proportional to $\gamma_\mu$.
Then the resulting
perfect lattice vertex has the same Dirac structure as the continuum vertex.
Hence only the left-handed lattice fermion couples to the lattice gauge
field, and gauge invariance is manifest in the vertex function. For $a>0$,
on the other hand, also the right-handed lattice fermion couples,
because now $a$ enters as an additional term in $\Delta^f$.
Then gauge
invariance is explicitly broken in the lattice action. Still it is guaranteed
that the theory is physically equivalent to the chiral gauge theory in the
continuum, because it was obtained analytically by a renormalization group
transformation. In particular the spectrum of the lattice theory does respect
gauge invariance.

\section{The exact axial current and the anomaly in the Schwinger model}

In this section we consider the $d=2$ Schwinger model with an abelian gauge
field $a_\mu$. We construct an exact lattice axial current and show that it
has the correct anomaly \cite{Bie95}.
For $a=0$ the renormalization group transformation and hence the perfect
action is chirally invariant. In contrast to the continuum the fermionic
measure on the lattice is also chirally invariant. This enables the formal
construction of a conserved axial Noether current $\tilde J^5_\mu$, and one
may argue that the anomaly vanishes on the lattice. In fact, assuming a
general form of $\Delta^f$ and $V_\mu$, Pelissetto showed that
$\langle \hat p_\mu \tilde J^5_\mu(p) \rangle_A = 0$ in an arbitrary
background lattice gauge field $A_\mu$, and he concluded that the anomaly is
cancelled by spurious ghost states \cite{Pel88}. At $a=0$, however, the
axial current $\tilde J^5_\mu$ is nonlocal, and there is no evidence for it
being related to the local current
\begin{equation}
\label{current}
j_\mu(x) = \bar\psi(x) \gamma_\mu \gamma_5 \psi(x)
\end{equation}
in the continuum, which has the anomaly
\begin{equation}
\langle p_\mu j^5_\mu(p) \rangle_a = \frac{e}{\pi} \epsilon_{\mu\nu} p_\mu
a_\nu.
\end{equation}
In fact, even in the continuum one can formally construct a gauge invariant
axial current
\begin{equation}
\tilde j^5_\mu(p) = j^5_\mu(p) - \frac{e}{\pi} \epsilon_{\mu\nu}
[a_\nu(p) - \frac{p_\nu p_\rho}{p^2} a_\rho(p)],
\end{equation}
which is anomaly free but nonlocal. Hence one must decide carefully what
lattice current to consider.

We now construct a lattice axial current that is related to the
continuum current $j^5_\mu$ by the renormalization group transformation.
A natural
definition is to consider the total flux of the continuum current through a
$(d-1)$-dimensional face $f_{\mu,x}$ that separates two hypercubes
$c_{x-\hat\mu/2}$ and $c_{x+\hat\mu/2}$. The corresponding lattice current
$J^5_{\mu,x}$ associated with the link connecting the lattice points
$x-\hat\mu/2$ and $x+\hat\mu/2$ is given by
\begin{equation}
J^5_{\mu,x} = \int_{f_{\mu,x}} dy \ j^5_\mu(y).
\end{equation}
We call this the `exact lattice axial current'.
Its standard lattice divergence
is by Gauss' law identical with the divergence of the continuum current
integrated over a hypercube $c_x$
\begin{eqnarray}
\delta J^5_x&=&\sum_\mu(J^5_{\mu,x+\hat\mu/2} - J^5_{\mu,x-\hat\mu/2})
\nonumber \\
&=&\int_{c_x} d^2y \ \partial_\mu j^5_\mu(y).
\end{eqnarray}
In momentum space it reads
\begin{eqnarray}
\hspace{-7mm}
&&J^5_\mu(p) = \sum_{l \in Z^2} j^5_\mu(p+2 \pi l) \Pi_{\neg\mu}
(p+2 \pi l) (-1)^{l_\mu}, \nonumber \\
\hspace{-7mm}
&&\Pi_{\neg\mu}(p) = \frac{p_\mu}{\hat p_\mu} \Pi(p).
\end{eqnarray}

We couple the current $J^5_\mu$ to an external lattice axial current $J^e_\mu$
by adding a term $(2 \pi)^{-2} \int_B d^2p \ J^e_\mu(-p) J^5_\mu(p)$ to the
continuum action and again we integrate out the continuum fields. Taking a
functional derivative of the lattice action with respect to $J^e_\mu(-p)$
one identifies
\begin{eqnarray}
J^5_{\mu}(p)&=&\frac{1}{(2 \pi)^2} \int_B d^2q \
\bar\Psi(p-q) \Delta^f(p-q)^{-1} \nonumber \\
&\times&\sum_{l,m \in \Z^2} \delta^f(p + 2 \pi l - q - 2 \pi m) \gamma_\mu
\gamma_5 \nonumber \\
&\times&\delta^f(q + 2 \pi m) \Delta^f(q)^{-1} \Psi(q) \nonumber \\
&\times&\Pi(p + 2 \pi l - q - 2 \pi m) \Pi(q + 2 \pi m) \nonumber \\
&\times&\Pi_{\neg\mu}(p + 2 \pi l) (-1)^{l_\mu} + O(e).
\end{eqnarray}
For reasons of space the $O(e)$ term can not be displayed here.

To show that the current has the correct anomaly we compute $\langle
J_\mu^5 (p) \rangle_A$ by integrating out the lattice fermions
\begin{eqnarray}
\hspace{-7mm}
&&\langle J_\mu^5 (p) \rangle_A = \int {\cal D}a \sum_{l \in \Z^2}
\langle j_{\mu}^{5}(p+2\pi l) \rangle_a \nonumber \\
\hspace{-7mm}
&&\times \Pi_{\neg\mu}(p+2\pi l) (-1)^{l_{\mu}} \exp(-s[a]) \nonumber \\
\hspace{-7mm}
&&\times \int {\cal D} D \exp\Big\{\frac{i}{(2 \pi)^2} \int d^2q
\nonumber \\
\hspace{-7mm}
&&\times D_\nu(-q) a_\nu(q) \Pi_\nu(q) \Big\} \nonumber \\
\hspace{-7mm}
&&\times \exp\Big\{- \frac{1}{(2 \pi)^2} \int_B d^2q \Big[i D_\nu(-q) A_\nu(q)
\nonumber \\
\hspace{-7mm}
&&+ \frac{1}{2} D_\nu(-q)(\alpha + \gamma \hat q_\nu^2) D_\nu(q)
\Big]\Big\}.
\end{eqnarray}
We insert the continuum quantity
$\langle j_\mu^5 (p) \rangle_a = (e/\pi )(p_\mu /p^2 )
\epsilon_{\nu \rho} p_\nu a_\rho (p)$,
which is known e.g. from Pauli-Villars regularization.
Integrating out also the continuum gauge field $a_\mu$ we arrive at
\begin{eqnarray}
\hspace{-7mm}
&&\langle J_\mu^5(p)\rangle = \frac{e}{\pi} \sum_{l \in \Z^2}
\frac{p_\mu + 2\pi l_\mu }{(p+2\pi l)^2} \epsilon_{\nu \rho}
\frac{p_\nu +2\pi l_\nu}{(p+2\pi l)^2} \nonumber \\
\hspace{-7mm}
&&\times \Pi_{\neg \mu}(p+2\pi l)
\Pi_\rho(p+2\pi l) (-1)^{l_\mu + l_\rho} \nonumber \\
\hspace{-7mm}
&&\times\Delta^g_\rho(p)^{-1} A_\rho(p).
\end{eqnarray}
Using $\hat p_\mu \Pi_{\neg\mu}(p + 2 \pi l)(-1)^{l_\mu} =
(p_\mu + 2 \pi l_\mu ) \Pi(p + 2 \pi l)$,
for the lattice divergence of the axial current one obtains
\begin{eqnarray}
\hspace{-7mm} &&
\langle \hat p_\mu J^5_\mu (p) \rangle_A = \frac{e}{\pi}
\sum_{l \in \Z^2} \epsilon_{\nu \rho} \frac{p_\nu + 2\pi l_\nu}
{(p+2\pi l)^2} \Pi (p+2\pi l)
\nonumber \\ \hspace{-7mm} &&
\times \Pi_\rho (p+2\pi l) (-1)^{l_\rho}
\Delta^g_\rho(p)^{-1} A_\rho(p). \label{deraxcur}
\end{eqnarray}
Finally we also construct the `exact lattice topological charge density'
\begin{equation}
Q_x = \frac{1}{\pi} \int_{c_x} d^2y \ \epsilon_{\nu \rho} \partial_\nu
a_\rho(y),
\end{equation}
by integrating the continuum charge density over a hypercube $c_x$. In momentum
space this implies
\begin{eqnarray}
Q(p)&=&\frac{1}{\pi} \sum_{l \in \Z^2} \epsilon_{\nu \rho}
(p_\nu + 2 \pi l_\nu) \nonumber \\
&\times&a_\rho(p + 2 \pi l) \Pi(p + 2 \pi l) (-1)^{l_\rho}.
\end{eqnarray}
Using eq.(\ref{aequation}) one sees that the exact
lattice axial current given in eq.(\ref{deraxcur}) coincides with
the topological charge density. In coordinate space this implies
\begin{equation}
\langle \delta J^5_x \rangle = e Q_x,
\end{equation}
which is the desired anomaly equation on the lattice.

\section{A nonperturbative version of the renormalization group
transformation}

To derive a perfect action for QCD beyond perturbation theory we want to
follow the lines of \cite{Has94,DeG95} and work with a renormalization
group transformation with a blocking factor 2. Our
goal is to formulate a nonperturbative version of the renormalization group
transformation whose perfect action reduces to the one of blocking from the
continuum at least for weak fields. Then we know that the perfect action is
extremely local (in lower dimensions even ultralocal), and therefore well
suited for numerical simulations.

Here we use compact link variables $U_{\mu,x} \in SU(N)$, where $x_\mu$ is
now an integer. We rescale $eA_\mu \to A_\mu$. Then the parallel transporters
are related to the gauge potential by
$U_{\mu,x} = \exp(i A^a_{\mu,x+\hat\mu/2} \lambda^a)$.
Under a gauge transformation $g$ the link variables transform as
\begin{equation}
\label{Utransformation}
^gU_{\mu,x} = g_x U_{\mu,x} g_{x+\hat\mu}^+.
\end{equation}
In the renormalization group
transformation we couple a blocked link $U'_{\mu,x'}$ on the coarse
lattice to an average of $2^d$ parallel transporters on the fine lattice. The
blocked link connects the centers of two neighboring $2^d$ blocks on the fine
lattice. For each point $x \in x'$ in the first block we construct a parallel
transporter as a product of two consecutive links starting
at $x$ and ending at the corresponding lattice point $x+2\hat\mu$ in the
second block. Then we couple the parallel transporters to the blocked link
$U'_{\mu,x'}$ by constructing a traceless anti-Hermitean matrix
\begin{equation}
B_{\mu,x'} = \frac{1}{2^d} \sum_{x \in x'} \log[U'_{\mu,x'}
(U_{\mu,x} U_{\mu,x+\hat\mu})^+]
\end{equation}
for each blocked link. A gauge transformation $g'$ on the coarse lattice
\begin{equation}
^{g'}U'_{\mu,x'} = g'_{x'} U'_{\mu,x'} {g'_{x'+2\hat\mu}}^+
\end{equation}
induces a gauge transformation $g_x = g'_{x'}$ for $x \in x'$ on the fine
lattice. Hence the anti-Hermitean matrix transforms as
\begin{equation}
^{g'}B_{\mu,x'} = g'_{x'} B_{\mu,x'} {g'_{x'}}^+.
\end{equation}
The field $B_\mu$ is a charged vector in the adjoint representation, not a
lattice gauge field. The renormalization group transformation contains the
covariant second derivative
\begin{eqnarray}
\hspace{-7mm}
&&\Delta_\mu[U'] B_{\mu,x'}=2 B_{\mu,x'} - U'_{\mu,x'} B_{\mu,x'+2\hat\mu}
{U'_{\mu,x'}}^+ \nonumber \\
\hspace{-7mm}
&&-{U'_{\mu,x'-2\hat\mu}}^+ B_{\mu,x'-2\hat\mu} U'_{\mu,x'-2\hat\mu}
\end{eqnarray}
of $B_\mu $. Under a gauge
transformation $g'$ it transforms as
\begin{equation}
\Delta_\mu[^{g'}U'] ^{g'} B_{\mu,x'} = g'_{x'}
\Delta_\mu[U'] B_{\mu,x'} {g'_{x'}}^+.
\end{equation}
The effective action for the coarse variables now takes the form
\begin{eqnarray}
\label{blockRGT}
\hspace{-7mm}
&&\exp(- S'[\bar\Psi',\Psi',U'])=\int {\cal D}\bar\Psi {\cal D}\Psi {\cal D}U
\nonumber \\
\hspace{-7mm}
&&\times\exp(- S[\bar\Psi,\Psi,U]) \ f[U] \nonumber \\
\hspace{-7mm}
&&\times\exp\Big\{- \sum_{x'} \frac{1}{a_2}
(\bar\Psi'_{x'} - \frac{b_2}{2^d}
\sum_{x \in x'} \bar\Psi_x) \nonumber \\
\hspace{-7mm}
&&\times(\Psi'_{x'} - \frac{b_2}{2^d}
\sum_{x \in x'} \Psi_x) - \frac{1}{2e^2} \sum_{x',\mu}
\nonumber \\ \hspace{-7mm}
&&\times
\mbox{Tr} \Big[B_{\mu,x'}^+
(\alpha_2 + \gamma_2 \Delta_\mu[U'])^{-1} B_{\mu,x'}\Big]\Big\}.
\end{eqnarray}
The functional $f[U]$ ensures the proper normalization of the partition
function and is given by
\begin{eqnarray}
\hspace{-7mm}
&&f[U]^{-1}=\int {\cal D}U' \exp\Big\{- \frac{1}{2e^2} \sum_{x',\mu}
\nonumber \\ \hspace{-7mm}
&&\times \mbox{Tr} \Big[B_{\mu,x'}^+
(\alpha_2 + \gamma_2 \Delta_\mu[U'])^{-1} B_{\mu,x'}\Big]\Big\}.
\end{eqnarray}
The parameters
\begin{eqnarray}
\hspace{-7mm}
&&a_2 = \frac{e^m - 1 - m}{2 m^2}, \,\,\,
b_2 = 2^{(d-1)/2}, \nonumber \\
\hspace{-7mm}
&&\alpha_2 = \frac{1}{8}, \,\,\,
\gamma_2 = - \frac{1}{128}
\end{eqnarray}
give an optimally local fixed point action.

It may seem that the above transformation is not gauge covariant, because
the fine lattice quark field $\Psi_x$ is not properly parallel transported to
the block center $x'$. However,
using gauge invariance of the fine action and of
the measure one realizes that the coarse action is indeed gauge invariant. The
above transformation corresponds to $K_\mu = 0$.

In the classical limit $e \rightarrow 0$ the full
renormalization group transformation reduces to the quadratic problem of
section 3 even for arbitrarily strong abelian gauge fields.
For 2-d configurations the fixed point action is ultralocal at the
quadratic level. Hence for any 2-d abelian gauge field the single plaquette
Manton action is classically perfect.

\section{Conclusions}

We have used the technique of blocking from the continuum to derive perfect
actions for fermions and gauge fields. Depending on the choice of the
renormalization group transformation the action is chirally invariant and
nonlocal, or local and chirally variant. In a chiral gauge theory this
corresponds to lattice formulations with unbroken or explicitly broken gauge
invariance. Since we always keep contact with the continuum theory in both
cases it is guaranteed that the spectrum of the lattice theory is identical
with the one of the continuum.

Also the axial anomaly is correctly reproduced as we have shown in the
Schwinger model. By construction our method of blocking from the continuum
extends to all orders of perturbation theory. Our construction does, however,
not yet provide a nonperturbative definition of lattice chiral fermions.
Still, we think that our analytic perturbative study provides insight into
this problem, and may inspire nonperturbative extensions.

The other aspect of our analysis concerns the perfect action for QCD. Our
perturbative results have led to an optimization of the parameters in the
renormalization group transformation. The perfect vertex function will be
important in the nonperturbative numerical determination of the perfect action
that is presently in progress \cite{BBW95}. This would apply to full QCD
with light quarks. The perturbatively perfect action for quarks
and gluons presented here may already be useful for simulations of
heavy quarks, where the leading cut-off effects are proportional to
the fermion mass. With our action these effects are eliminated up to $O(e)$.


We are indebted to R. Brower, T. DeGrand, P. Hasenfratz,
A. Kronfeld, H. Neuberger, F. Niedermayer and Y. Shamir for many stimulating
discussions.


\begin{thebibliography}{22}

\bibitem{Kog74} K. Wilson and J. Kogut, Phys. Rep. C12 (1974) 75.

\bibitem{Wil76}
K. Wilson, in New Pathways in High Energy Physics II, ed. A. Perlmutter
(Plenum, New York 1976) 243.

\bibitem{Bie95}
W. Bietenholz and U.-J. Wiese, Preprint MIT-CTP 2423 (hep-lat/9503022).

\bibitem{Has94}
P. Hasenfratz and F. Niedermayer, Nucl. Phys. B 414 (1994) 785.

\bibitem{DeG95}
T. DeGrand, A. Hasenfratz, P. Hasenfratz and F. Niedermayer, Preprint
BUTP-95/14 (hep-lat/9506030), BUTP-95/15 (hep-lat/ 9506031),
BUTP-95/30 (hep-lat/9508024).

\bibitem{Lep93}
G. P. Lepage and P. Mackenzie, Phys. Rev. D48 (1993) 2250.

\bibitem{Wie93}
U.-J. Wiese, Phys. Lett. B315 (1993) 417.

\bibitem{BW95}
W. Bietenholz and U.-J. Wiese, in preparation.

\bibitem{Pel88} A. Pelissetto, Ann. Phys. 182 (1988) 177.

\bibitem{BBW95}
W. Bietenholz, R. Brower and U.-J. Wiese, in preparation.

\end{thebibliography}
\end{document}